\documentclass[aps,pre,reprint,superscriptaddress,amsmath,showpacs]{revtex4-1}

\usepackage[breaklinks,colorlinks=true]{hyperref}
\usepackage{graphicx}

\begin{document}

\title{Flocking at the edge of chaos}

\author{Kunal Bhattacharya}
\email[Corresponding author; ]{kunal.bhattacharya@aalto.fi}
\affiliation {Department of Information and Computer Science, Aalto University, P.O. Box 15400, FI-00076, Finland}

\author{Abhijit Chakraborty}
\affiliation {The Institute of Mathematical Sciences, CIT Campus - Taramani, Chennai 600113, India}

\date{\today}

\begin{abstract}
Recent investigations have provided important insights into the complex structure and dynamics of collectively moving flocks of living organisms. Two intriguing observations are -- scale-free correlations in the velocity fluctuations, in the presence of a high degree of order, and topological distance mediated interactions. Understanding these features, especially, the origin of fluctuations, appears to be challenging in the current scheme of models. It has been argued that flocks are poised at criticality. We present a self-propelled particle model where neighbourhoods and forces are defined through topology based rules. The force fluctuations occur spontaneously, and gives rise to scale-free correlations in the absence of noise and in the presence of alignment of velocities. We characterize the behaviour of the model through power spectral densities and the Lyapunov spectrum. Our investigations suggest self-organized criticality as a probable route to the existence of criticality in flocks.  
\end{abstract}


\maketitle

\begin{figure}
\centering
\includegraphics[width=0.75\columnwidth]{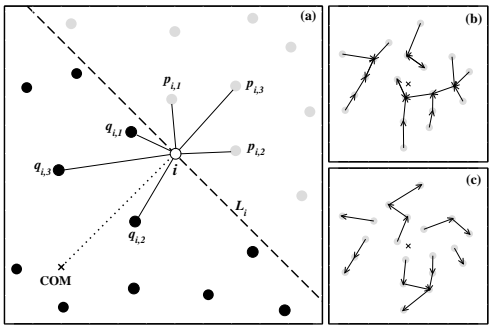}
\caption{\label{fig:mod_net}(a) The figure illustrates the choice of neighbours for a particle $i$ with $m=3$ at a certain instant of time. The dotted line connects the location of $i$ to the centre of mass of the system (COM, denoted by a cross). The dashed line $L_i$ is perpendicular to the dotted line and passes through the location of $i$. The line $L_i$ divides the two dimensional space into two non-overlapping regions. We first consider the region in which the COM is located. We consider the particles lying in this region as {\it particles located in the direction of the COM}. (In general, particles lying on $L_i$ are also included in this region). Such particles are shown as filled black circles. In this region, the particles $q_{i,1}$, $q_{i,2}$ and $q_{i,3}$ are the first three ($m=3$) nearest neighbours of $i$ and constitute the set $Q_i$. The particles in the other region (denoted by filled grey circles) are considered as  {\it particles located in the direction opposite to the direction of the COM}. The set of neighbours in this region is $P_i=\{p_{i,1},p_{i,2},p_{i,3}\}$. Together the sets $Q_i$ and $P_i$ constitute the neighbourhood of $i$ and are updated at each instant. The figures (b) and (c) schematically shows the sets $Q$ and $P$ for individual particles when $m=1$ in a system of $20$ particles at a certain instant. In the figure (b) an arrow points from each particle to its neighbours in the set $Q$. In (c) the arrow connects to neighbours in $P$. Note, that for some of the particles there are no neighbours in the set $P$.}
\end{figure}

The last decade has seen a growing interest in the study of collective motion of  living as well as non-living systems. Examples include, collectively swimming bacteria \cite{Sokolov07}, human moshing \cite{Silverberg13}, propulsion of Janus particles \cite{Jiang10} and motion in granular systems \cite{Deseigne10}. Particularly worth mentioning are the investigations of the StarFlag project on the structure and dynamics of three dimensional starling flocks. It was found that any bird in a flock of starlings, on the average, interacts with a fixed number of its nearest neighbours \cite{Ballerini08}. This notion of `topological distance', for defining the neighbourhood, is contrary to the assumptions in models \cite{Vicsek12} that a bird would interact with all other birds within a fixed metric distance. Moreover, the analyses revealed the existence of scale-free correlations in the velocity fluctuations of the birds \cite{Cavagna10}. Such scale-free correlations were also found in motile bacterial colonies \cite{Chen12} and midge swarms \cite{Attanasi14}. Interestingly, in case of starlings and bacteria the flocks showed a degree of order. The velocities of the individuals were found to be aligned, on the average. 

Critical fluctuations in models are generally obtained by tuning control parameters like noise and fields. However, the mechanism through which the fine tuning of the parameters transpires in real flocks is not yet known. It has been argued that flocks are actually poised at criticality \cite{Mora11}. Maximum entropy modeling based on data \cite{Bialek12}, has been used to extract the values of the parameters that make a flock critical. The contrasting paradigm of self-organized criticality (SOC) has also been suggested as a dynamical route to the realization of criticality \cite{Cavagna10,Mora11}. In this paper, we provide a self-propelled particle (SPP) model along the latter lines. In our model, the critical state is realized in the absence of any external noise or field. We implement a topological distance based rule for the choice of the neighbourhood of the particles. We assume that particles in the flock interact with a certain fixed number nearest-neighbours through spring-like forces. We assume that particles have a preference to get attracted towards the neighbours that are located towards the centre of mass (COM). Previously, one of the authors considered a comoving boundary, defined with respect to the COM, where, the particles at or outside the boundary would be attracted towards the COM \cite{Bhattacharya10}. In a recent investigation, employing Vornoi tessellation, the particles at the convex hulls of the surface were considered to get attracted towards bulk \cite{Pearce14}.       

We consider $N$ identical particles on the two dimensional space characterized by the positions, \{$\vec{r}_i$\}, and velocities \{$\vec{v}_i$\}, where, $1\leq i\leq N$. We assume that the particles interact with each other through forces depending on topological distances. At every instant, the motion of a particle is influenced by its $2m$ ``neighbours'', where $m$ is a parameter in the model. To decide the instantaneous set of neighbours for particle $i$ we divide the entire space into two non-overlapping regions, using the position of $i$ and the location of the COM for the entire set of $N$ particles. We consider the particles lying in the region, in which the COM lies, as particles located in the direction of the COM, with respect to the particle $i$. The $m$ nearest particles in this region constitute the set $Q_i$. Similarly, the $m$ nearest particles in the other region make up the set $P_i$. The sets, $P_i$ and $Q_i$, together comprise the neighbourhood of $i$. This rule is illustrated in Fig. \ref{fig:mod_net}.  


At every instant of time the sets $Q$ and $P$ for each of the particles are constructed and the positions and velocities are calculated using the equations of motion: $\partial \vec{r}_i/\partial t=\vec{v}_i$, and, $M\,\partial \vec{v}_i/\partial t=\vec{F}(\vec{v}_i)+\vec{F}_{Q,i} +\vec{F}_{P,i}$, where, $M$ is the mass of each particle. The propulsion force, $\vec{F}(\vec{v}_i)=-b\left({|\vec{v}_i^2|}/{v_0^2}-1\right)\vec{v}_i$, is the Rayleigh-Helmholtz (RH) friction \cite{Klimontovich97,Ebeling04,D’Orsogna06}, 
where, $v_0$ is the optimal speed, which makes the propulsion force zero; and the parameter $b$ controls the strength of the propulsion force relative to the other forces. The term $\vec{F}_{Q,i}$ and $\vec{F}_{P,i}$ are the forces acting on $i$ due to the neighbours in the sets $Q_i$ and $P_i$, respectively. We assume that the force of interaction of $i$ with a neighbour $j$ in the set $Q_i$ is essentially spring-like with $\vec{F}_{Q,i}=-(1/n_{Q,i})\sum_{j\in Q_i}k(|\vec{r}_{ij}|-a)\vec{r}_{ij}/|\vec{r}_{ij}|$, where, $\vec{r}_{ij}=\vec{r}_i-\vec{r}_j$, $k$ is the spring constant, $a$ is the optimal separation with a neighbour, and $n_{Q,i}$ is the number of neighbours in the set $Q_i$. However, for the neighbours in $P_i$ we consider only repulsive interactions below the separation $a$, such that, $\vec{F}_{P,i}=-(1/n_{P,i})\sum_{j\in P_i}f_{ij}k(|\vec{r}_{ij}|-a)\vec{r}_{ij}/|\vec{r}_{ij}|$, where, $f_{ij}=1$ when $|\vec{r}_i-\vec{r}_j|<a$ and $f_{ij}=0$, otherwise. We note here that, the condition $n_{Q,i}=m$ is always satisfied when $m\ll N$, but this is not the case with $n_{P,i}$. These can be seen in Fig. \ref{fig:mod_net}(b) and (c). For such particles the total number of neighbours is less than $2m$. 

We study the time evolution of the model by numerically integrating the equations  of motion using the velocity-Verlet algorithm \cite{modif-Verlet}. We set the scales for measurement as $M$, $a$ and $v_0$. All quantities are measured in these scales. For our main investigation, we take $b=1.0$ and $k=3.0$, and the integration time step $dt=0.005$. At every time step we update the set of neighbours for every particle. However, in the absence of a cut-off distance for the cohesive force, the construction of sets $P$ and $Q$ becomes computationally intensive. 


At the beginning of the simulations we randomly position all the particles in circular a region of having radius $\sqrt{N}/2$ and with randomly oriented velocities having magnitude unity. We simulate the model in open boundary conditions and, in general,  we find that the flock is a single cohesive entity. On the average the particles remain  bounded in a finite region when in the frame of the COM. This is ensured by cohesive forces on the particles in the direction of the COM as well as the absence of any external noise. Although, artificial initial conditions for positions and velocities can be constructed which may make the flock to expand in space as time progresses, we did not encounter such cases, starting from arbitrary initial conditions. 
      
\begin{figure}
\centering
\includegraphics[width=0.75\columnwidth]{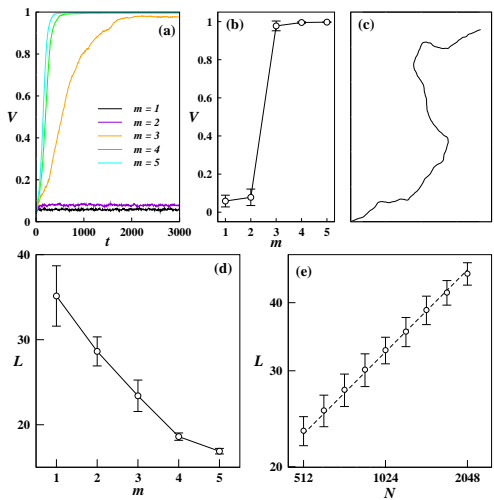}
\caption{\label{fig:param_plots}(Color online) (a) The time evolution of the configuration averaged speed of the centre of mass ($V$) is shown for different values of $m$. (b) The value of $V$ in the steady state (as seen from the figure (a)) is plotted against $m$. (c) A typical trajectory of the centre of mass for a flock with $m=3$, monitored over $2000$ time steps. (d) The linear size of the flock ($L$) in the steady state is plotted against $m$.  For (a)--(d), $N=512$. The straight lines in (b) and (d) are guides to the eye. (e) Log-log plot of the linear size of the flock ($L$) plotted against number of particles ($N$) when $m=3$. The dashed line is suggestive of $L$ increasing as $N^{\zeta_1}$ with $\zeta_1=0.5$. The bars indicate standard deviation.}
\end{figure}

\begin{figure}
\centering
\includegraphics[width=0.75\columnwidth]{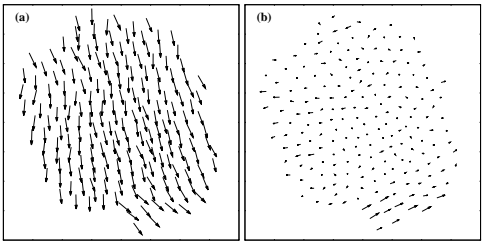}
\caption{\label{fig:param_config}(a) A typical flock in the steady state generated by the model corresponding to $N=200$ and $m=3$. The instantaneous velocity vectors of the particles are plotted. (b) The fluctuations in velocities  (with respect to the velocity of the COM) corresponding to the figure (a) are plotted.}
\end{figure}

The velocity of the COM for the $N$ particles is, $\vec{V}=\frac{1}{N}\sum_i \vec{v}_i$. We characterize the alignment of velocities using the speed, $V=|\vec{V}|$. We would like to mention that, having attractive interactions with neighbours in $P$, causes the flock to stretch and with $V\simeq0$ producing weak translation of the COM. We exclude such interactions in our model. In Fig. \ref{fig:param_plots}(a) we show the approach to the steady state in our model, for five different values $m$. The values of $V$ in the steady state is plotted in Fig. \ref{fig:param_plots}(b). In general, for values of $m\geq 3$ the model produces flocks having $V\simeq 1.0$. The case $m=3$ is shown in Fig. \ref{fig:param_config}(a). For $m<3$, the flocks fail to show any global order. Remarkably, in the case $m=3$, not only the value of  $V$ is high, the fluctuations in $V$ are high as well (from Fig. \ref{fig:param_plots} (b), $V=0.977\pm0.025$). This also results in the tortuous trajectory for the COM (Fig. \ref{fig:param_plots}(c)). This is in contrast with the cases, $m=4$ and $5$, where the fluctuations appear to be rather suppressed (for $m=5$, $V=0.997\pm0.002$).  This is the main motivation for choosing $m=3$ for the investigation of correlations in velocity fluctuations. 

The fluctuation in the velocity of a particle $i$, is defined as, $\vec{u}_i=\vec{v}_i-\vec{V}$. In Fig. \ref{fig:param_config}(b) we plot the fluctuations in the velocities for the particles when $m=3$. We define the linear size ($L$) of the flock in the following fashion. At any instant we find the size by calculating the maximum of all distances $|\vec{r}_i-\vec{r}_j|$, where, $i,j\in[1,N]$ \citep{Cavagna10,Chen12}. We find $L$ by averaging the size over initial conditions and time snapshots. In Fig. \ref{fig:param_plots}(d) we plot $L$ for different values of $m$. The plot reveals larger variation in the shape of the flock for smaller values of $m$.
For, $m=3$, we consider the variation in the size of the flock with the number of particles in system. In Fig. \ref{fig:param_plots}(e) we plot $L$ versus $N$. We find that $L\sim N^{{\zeta}_1}$, with $\zeta_1=0.5$. For bacterial clusters  moving in two dimensions this exponent was found to be $0.6$ \cite{Chen12}. Starling flocks have three-dimensional structure and have been found to be asymmetrical. Interestingly, a regression fit, to the data provided \cite{Cavagna10} for the number of birds in flocks and the corresponding sizes, results in the exponent $0.6$. 

The spatial correlation function for velocity fluctuations \cite{Cavagna10,Cavagna13} is defined as, $C(r)=(1/C_0){\sum_{i,j}\vec{u}_i\cdot\vec{u}_j\,\delta (r-r_{ij})}/{\sum_{i,j}\delta(r-r_{ij})}$, where, $C_0$ is a normalization constant, such that, $C(0)=1$. The Dirac $\delta$-function is realized through the binning of the correlation function. The correlation length, $\xi$, is defined through the relation $C(r=\xi)=0$, and is expected to provide the average size of the correlated domains. For starling flocks \cite{Cavagna10} and bacterial clusters \cite{Chen12}, the correlation length is found to be proportional to the linear size ($L$) of the system. Absence of a characteristic scale for domains implies scale-free correlations. A finite-size scaling ansatz for the correlation function shows that the correlation function evaluated in the limit of $L\to\infty$, $C_{\infty}(r)\sim r^{-\gamma}$. The exponent is estimated from the relation $C'(r/\xi=1)\sim -L^{-\gamma}$ \cite{Cavagna10}. Starling flocks revealed a extremely long-ranged correlations ($\gamma\sim 0$). Models have shown the existence of scale-free correlations in the highly ordered regime \cite{Cavagna13,Niizato12,Huepe14} mainly through the relation, $\xi(L)\propto L$. A $2d$ SPP model \cite{Ginelli10} using topology (Voronoi cells) to determine nearest neighbours, studied in the low-noise regime, yielded $\gamma=0.4$ \cite{Cavagna13}. The $3d$ classical Heisenberg model with a dynamical external field and at low noise, showed $\gamma$ is $1$ and $0$, in weak and strong fields, respectively \cite{Cavagna13}. 

\begin{figure}
\centering
\includegraphics[width=0.75\columnwidth]{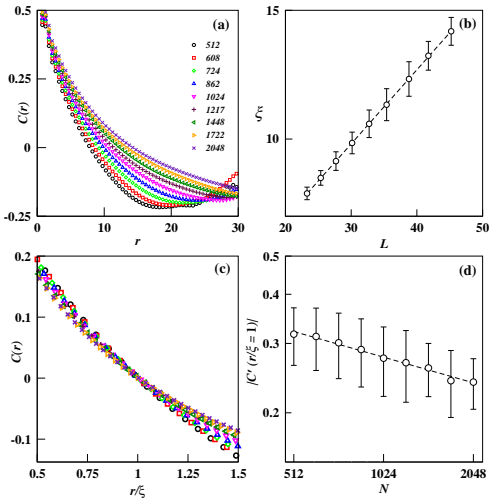}
\caption{\label{fig:corr_plot} (Color online) (a) Plot of the correlation functions $C(r)$ for velocity fluctuations for flocks with different $N$. (b) Plot of correlation length $\xi$ versus the linear size $L$. The dashed line is a fit of the form $\xi=c_0+c_1L$, giving $c_0=1.17$ and $c_1=0.29$. (c) $C(r)$ plotted  against rescaled separation $r/\xi$. The symbols denoting different values of $N$ in (a) and (c) are provided in the legend of (a). (d) Log-log plot of the slopes of the  correlation functions plotted in (c) at $r/\xi=1$, as a function of $N$. The dashed line is a power-law fit having the form $|C'(r/\xi=1)|\sim N^{-\zeta_2}$ giving $\zeta_2=0.22$. The bars in (b) and (d) indicate the standard deviations obtained from averaging over $100$ time snapshots (separated by $10^4$ time steps) in the steady state starting from around $100$ different initial conditions.} 
\end{figure}

We plot $C(r)$ in Fig. \ref{fig:corr_plot}(a) for nine different values of $N$. We calculate $\xi$ by averaging the positions of zero-crossings of $C(r)$. The Fig. \ref{fig:corr_plot}(b) suggests that the correlations are indeed scale-free. We find that, the general behaviour for large values of $L$ is, $\xi(L)\approx c_1.L$, with, $c_1=0.29$. The values of $c_1$ in the cases of starling flocks and bacterial clusters are $0.35$ and $0.3$, respectively. In Fig. \ref{fig:corr_plot}(c) we plot the correlation functions against rescaled length $r/\xi$. The figure suggests a decay of the slope calculated a $r/\xi=1$ with increase in $N$. We determined the slopes, $|C'(r/\xi=1)|$ during the averaging of $C(r)$ and $\xi$. The absolute values of the slopes are plotted with $N$ in Fig. \ref{fig:corr_plot}(d). A power-law fit suggests $|C'(r/\xi=1)|\sim N^{-\zeta_2}$, with $\zeta_2=0.22$. Combining this result with dependence of $L$ on $N$, we find, $\gamma=\zeta_2/\zeta_1=0.44$. This value suggests that in our model the decay of correlations is faster compared to starling flocks but is very similar to that observed in the low-noise regime of the topological SPP model \cite{Cavagna13,Ginelli10}. We would like to mention here that we also investigated the correlations in the fluctuation of speed \cite{Cavagna10,Chen12}. We find the overall behaviour to be qualitatively similar to that of correlations in velocity fluctuations. However, the absolute values for correlations and correlation lengths appear to be lesser. A possible reason for this would be the RH friction which suppresses the fluctuations in speed. 

\begin{figure}
\centering
\includegraphics[width=0.75\columnwidth]{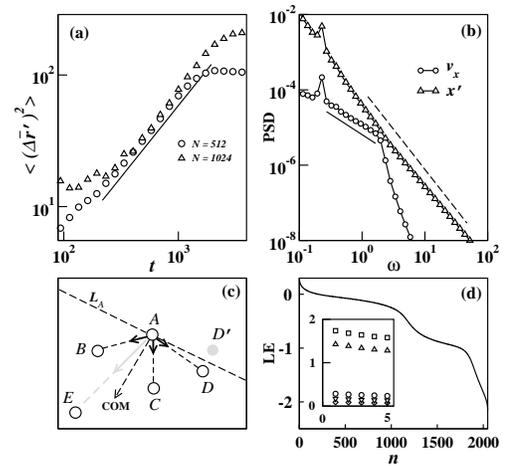}
\caption{\label{fig:dynam_plot}(a) Log-log plot of the configuration averaged mean square displacement in the COM frame for tagged particles against time. The different symbols correspond to different flock sizes as indicated in the legend. The straight line, having slope $\beta=1.09$, is a guide for the eye. Here, $\beta$ is obtained by averaging the slopes of the individual curves. (b) Log-log plot of the power spectral density (PSD) against the angular frequency ($\omega$) corresponding to dynamical variables of a tagged particle ($N=512$): (i) the $x$-component of the velocity ($v_x(t)$), and (ii) the $x$-component of position in the COM frame ($x'(t)$). The lengths of the time series' in both cases are equal to $2^{18}$ time steps. The straight lines indicate power-law fits for the PSD curves. The solid straight line has a negative slope, $\alpha_v=1.08$ and the dashed line has a negative slope, $\alpha_x=2.21$. (c) Illustration of the mechanism through which fluctuations in forces occur. Refer to main text for details. (d) Lyapunov spectrum showing the $4N$ exponents with $N=512$. The spectrum is obtained after a convergence time $10^4$ time steps. The inset plot shows the five largest Lyapunov exponets for values of $m=1$ (square), $2$ (triangle up), $3$ (circle), $4$ (triangle down) and $5$ (diamond).}
\end{figure}

We also study the diffusion of particles in the COM frame. For starling flocks the diffusive behaviour reflects the fact that the set of neighbours for a given bird keeps changing in time \cite{Cavagna13a}. At any given instant we consider a particle to be at the boundary if the number of neighbours in the set $P$ is less than $m$. We tag individual particles and follow their motion until they reach the boundary. We calculate the average of the squared displacements, {\small $\langle(\Delta$}$\vec{r'}${\small $)^2\rangle$}, of these particles in the centre of mass frame.     The Fig. \ref{fig:dynam_plot}(a) reveals, {\small $\langle(\Delta$}$\vec{r'}${\small $(t))^2\rangle$}$\sim t^\beta$, with $\beta=1.09$. This indicates a normal diffusive behaviour in contrast to super-diffusive behaviour observed for starlings, where $\beta=1.73$. 

We investigated the nature of the spontaneous fluctuations by analyzing the time variation of different quantities. We calculated the power spectral density (PSD) corresponding to the time series' for the $x$-component of the velocity ($v_x$) of a particle and the $x$-component of its position ($x'$) in the COM frame. The results are shown in Fig. \ref{fig:dynam_plot}(b). The PSDs are averaged over several time series and plotted against frequency $\omega$. We find the PSD for $v_x$ has a power-law region of the form $\omega^{-\alpha_v}$, with, $\alpha_v=1.08$. The upper bound to the power-law approximately coincides with the natural frequency, corresponding to the spring-like interactions, that is $\sqrt{k/M}=\sqrt{3}$ (in units of $v_0/a$). The peak appearing in the lower frequency region can be be attributed to the direction reversals of the particle near the boundary of the flock. The PSD for $x'$ is a power-law over most of the frequency span and decays as $\omega^{-\alpha_x}$, with, $\alpha_x=2.21$. This signifies Brownian nature of the time series for $x'$ and confirms the diffusive behaviour seen from Fig. \ref{fig:dynam_plot}(a). We believe that the origin of scale-free correlations in the absence of noise suggests that the system is in a state of self-organized criticality (SOC). The $1/\omega^{\alpha_v}$ decay of the PSD for velocity reaffirms this idea.

The reason that the fluctuations persist in the system when $m\geq3$, even in the absence of noise, is mainly the absence of any stable structure in the COM frame. Consider the subsystem shown in Fig. \ref{fig:dynam_plot}(c) where $m=3$. The figure shows a particle $A$ and its neighbours in the set $Q_A$.  We assume that any at any instant $Q_A=\{B,C,D\}$ and the distances between $A$ and the neighbours is just greater than $a$, the optimal separation. Also, let all the velocities be aligned and of magnitude $v_0$. This configuration results in $A$, experiencing attractive forces (shown with solid black arrows), small in magnitude and directed towards individual particles. The resultant force $\vec{F}_{Q,A}$ is an average vector, and is expected to be small, as well. Now, consider a perturbation, displacing the particle $D$ by small amount such that its new position is above the line $L_A$. The configuration would result in $E$ replacing $D$ in the set $Q_A$. Due to the location of $E$, the new $\vec{F}_{Q,A}$ would be of larger magnitude and would have a different direction. In fact, the direction of force is expected to further increase the separation with $D$ and would not allow $D$ to get restored as neighbour. The above mechanism prevents the particles to settle down into a `frozen' structure in the COM frame, when beginning from arbitrary initial conditions. As $m$ increases, because of the presence of a larger number of neighbours, averaging of the force is better and the characteristic fluctuation is lesser. For $m<3$, certain stable structures may be retained when taken as initial conditions. However, such structures are not observed for arbitrary initial conditions.

The fact that perturbations are able to proliferate in the system, generally implies sensitivity to initial conditions. This prompted us to calculate the Lyapunov spectrum for the dynamics. We use the Gram--Schmidt orthonormalization procedure \cite{Shimada79,Wolf85}. The converged spectrum for $N=512$ and $m=3$ with $4N$ exponents is shown in Fig. \ref{fig:dynam_plot}(d). In the inset we compare this with the spectra for other values of $m$ by plotting the first five Lyapunov exponents. For all the values of $m$, the system has positive exponents and hence indicate exponential instability. For $m=1$ and $m=2$, the motion of particles remain confined to a region ($V\approx 0$), and the largest Lyapunov exponents (LLEs) are $1.72$ and $1.42$, respectively. This suggests that the motion is chaotic. However, for larger values of $m$ the motion is unbounded and LLEs are smaller ($0.28$, $0.17$, and $0.08$ for $m=3$, $4$ and $5$, respectively). We believe that for $m=3$ the flock is poised at the edge of chaos, which is generally assumed to be the case with SOC \cite{Bak97}. The nature of the dynamics is very similar to that of the train model \citep{Vieira96}, where the self-organized critical states were shown to have a positive LLE.

It  turns out that the cohesion between particles is a key ingredient for the development of long-ranged correlations. The calculation of the force, in case of individual particles, implicitly depends on the knowledge of the COM of the system. This rule allows the flock to be cohesive unit even in the presence of fluctuations and the flock is always in the zero-density limit. A simple topological rule may not be sufficient to prevent the fragmentation of the flock into smaller clusters and subsequent expansion in the space, when simulated in open boundary conditions. We would like to argue that the case in real flocks may not be completely different. Birds may actually detect the density of individuals over a region and might have a preference of joining regions of higher densities, and, in effect, getting pulled towards the COM. Bacterial aggregates are studied over a quasistationary state, where a finite density is guaranteed. Also, note, that in the steady state of the model, the actual forces appear locally on scales much smaller than the flock size.

In conclusion, we provide an SPP model with a mechanism through which critical fluctuations can appear inside flocks in the presence substantial amount of order and in the absence of external noise. Noise in the flock is rather internal,  where, force fluctuations are caused by reshuffling of neighbourhoods and vice versa. We have also included the concept of topological distance while defining neighbourhood of particles. In future, we plan to include the effect of noise and external fields in our model, and characterize the response of the flock. Also, it would be interesting to construct such a model in the evolutionary paradigm \cite{Hidalgo14} where the susceptible state naturally evolves as stable strategy.

\begin{acknowledgments}
KB and AC acknowledge helpful discussions with T. Vicsek and A.~K. Nandi. 
\end{acknowledgments}

\end{document}